\documentstyle[twocolumn,aps,graphicx,floats]{revtex}

\begin{document}

\draft

\title{Magnetic Field Dependent Tunneling in Glasses}

\author{P. Strehlow, M. Wohlfahrt}
\address{Physikalisch-Technische Bundesanstalt, Abbestra\ss{}e 2-12, D-10587 Berlin, Germany}

\author{A.G.M. Jansen}
\address{Grenoble High Magnetic Field Laboratory, MPI-CNRS, B.P. 166, F-38042 Grenoble Cedex 9, France}

\author{R. Haueisen, G. Weiss}
\address{Physikalisches Institut, Universit\"at Karlsruhe, D-76128 Karlsruhe, Germany}

\author{C. Enss and S. Hunklinger}
\address{Institut f\"ur Angewandte Physik, Universit\"at Heidelberg, Albert-Ueberle-Stra\ss{}e 3-5, D-69120 Heidelberg, Germany}

\date{\today}
\maketitle
\begin{abstract}
We report on experiments giving evidence for quantum effects of electromagnetic flux in barium alumosilicate glass. In contrast to expectation, below $100\,$mK the dielectric response becomes sensitive to magnetic fields. The experimental findings include both, the complete lifting of the dielectric saturation by weak magnetic fields and oscillations of the dielectric response in the low temperature resonant regime. As origin of these effects we suggest that the magnetic induction field violates the time reversal invariance leading to a flux periodicity in the energy levels of tunneling systems. At low temperatures, this effect is strongly enhanced by the interaction between tunneling systems and thus becomes measurable.
\end{abstract}


\narrowtext
The low-temperature properties of glasses have been attributed to low-energy excitations present in nearly all amorphous solids and disordered crystals \cite{esquinazi}. Considerable theoretical and experimental investigations have been expended to understand these excitations. In the standard tunneling model (TM) \cite{anderson} they are described on a phenomenological basis by non-interacting two-level tunneling systems (TLS). These TLSs are thought to consist of small atomic entities which are able to tunnel between at least two equilibrium positions. Thus, a TLS can be approximately treated like a particle moving in an asymmetric double-well potential. The excitation energy between the two lowest states of the asymmetric double-well, $E = \sqrt{{\it\Delta}^2 + {\it\Delta}_0^2}$, is determined by the asymmetry~${{\it\Delta}}$ and the tunnel splitting~${\it\Delta}_0$. Because of the random structure of glasses ${\it\Delta}$ and ${\it\Delta}_0$ are broadly distributed. According to the TM a distribution $P({\it\Delta},{\it\Delta}_{0}) = \overline{P}/{\it\Delta}_{0}$ is assumed, where $\overline{P}$ is a constant.

Treating the coupling of TLSs to external acoustic and electric fields as a weak perturbation, the TM successfully explains many of the anomalous thermal, acoustic and dielectric low-temperature properties of glasses. Various recent experiments, however, demonstrate considerable deviations from this theory. In particular, we refer here to the permittivity (dielectric constant)~$\varepsilon'$ which levels off at very low temperatures \cite{rogge} whereas the TM predicts a $\ln(1/T)$ behavior. Interaction between TLSs has been suggested as a possible origin for these discrepancies \cite{caruzzo}, although a comprehensive theory is still missing. Phenomenologically, a low energy cut-off~${\it\Delta}_{\rm 0,min}$ may be introduced for the spectral density of tunneling states to account for the levelling-off. Values of ${\it\Delta}_{\rm 0,min}/k_{\rm B}\approx 1\,{\rm mK}$ indeed describe the $\varepsilon'(T)$ data quite well. However, applying the TM to the dielectric and acoustic behavior of glasses at higher temperatures requires values of ${\it\Delta}_{\rm 0,min}/k_{\rm B}<10^{-3}\,{\rm mK}$. The remarkably enhanced magnitude of ${\it\Delta}_{\rm 0,min}$ at very low temperatures indicates that interaction between TLS leads to a renormalization of essential parameters of the TM such as $\overline{P}$ and ${\it\Delta}_0$. In this case quasiparticles are considered rather than bare non-interacting TSs. Such renormalization effects might partially account for the general success of the standard tunneling model although this has not been explicitly stated yet.

Moreover, it is conceivable that due to interaction between TLSs the action of electromagnetic fields on the quantum-mechanical state of charged TLSs becomes measurable. This is motivated mainly by the recently reported observation of the dielectric response of multicomponent glasses being extremely sensitive to weak magnetic fields \cite{strehlow1}. A theoretical interpretation as magnetic flux effect (Aharonov-Bohm effect) can be given in a generalized tunneling model \cite{kettemann}.

In the present paper we report on measurements of the dielectric response of a BaO-Al$_{2}$O$_{3}$-SiO$_{2}$-glass in magnetic fields ranging from a few mT up to $25\,$T. Indeed, we were able to demonstrate the existence of quantum effects of electromagnetic flux in glasses: magnetic fields cause drastic changes of the dielectric response and lead to oscillatory variations. 

Barium alumosilicate glass is characterized by a large intrinsic polarizability. Therefore, thick-film sensors based on this glass have already been used in glass capacitance thermometers \cite{strehlow2} and also in previous experiments at very low temperatures \cite{strehlow1}. In order to substantiate the existence of the surprising phenomena we are going to report, and to exclude experimental artifacts, the measurements were carried out in Berlin, Karlsruhe and Grenoble under different experimental conditions. In the Grenoble experiment at fields up to $25\,$T the sensor was inserted into the mixing chamber, whereas in Berlin it was mounted on a silver post bolted to a nuclear demagnetization stage. In Karls\-ruhe it was attached to a silver cold-finger connected to the mixing chamber of a dilution refrigerator. The measurements were performed at $1\,$kHz either with a home-made bridge, consisting of a nine-decade inductive voltage divider and a lock-in amplifier, or with a self-balancing capacitance bridge (Andeen-Hagerling, model~2500A). Based on an extended temperature scale \cite{schuster} the temperature was measured by means of a $^3$He melting curve thermometer, an Au:Er susceptibility thermometer, and a pulsed platinum NMR thermometer. In the three experiments with substantially different setups fully consistent results were obtained.

\begin{figure}[b]
\vbox{
\includegraphics{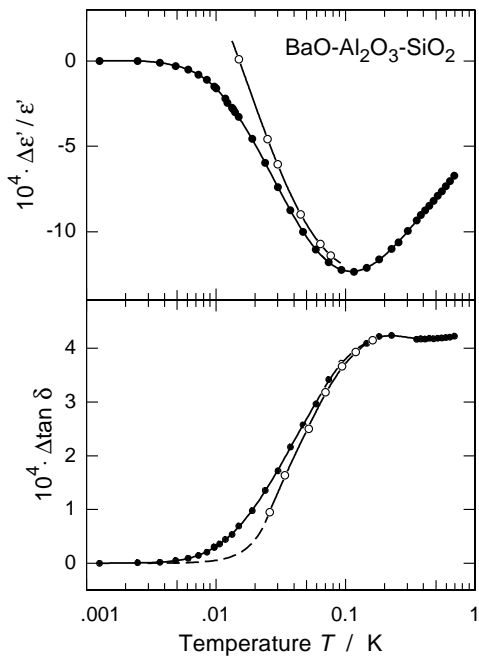}
\caption{Temperature variation of the permittivity $\Delta\varepsilon' / \varepsilon'$ and the dielectric loss $\Delta\tan\delta$ of  a $50\,\mu$m thick sample of BaO-Al$_2$O$_3$-SiO$_2$-glass measured at a frequency of $1\,$kHz and an excitation voltage~$U=0.75\,$V (full circles). In both cases the values at $T_{\rm 0}=1.26\,$mK were taken as a reference. The open circles represent the maximum value of $\varepsilon'$ and the minimum value of $\tan\delta$ in the magnetic field as discussed in the text. The dashed line shows the $T^3$-dependence of the dielectric loss as predicted by the TM.}
}%
\label{fig1}
\end{figure}

The temperature dependence of the permittivity~$\Delta\varepsilon' (T)/\varepsilon' = [\varepsilon'(T) -\varepsilon'(T_{\rm 0})] /\varepsilon'(T_{\rm 0})$ and of the dielectric loss $\Delta\tan\delta (T) =\tan\delta(T)- \tan\delta(T_{\rm 0})$ without magnetic field is shown in Fig.~1 by the full dots (open symbols refer to measurements at finite fields and will be discussed later). In qualitative agreement with the predictions of the TM, $\varepsilon'$ decreases logarithmically with decreasing temperature, passes a minimum at $T_{\rm m} = 113\,$mK and increases logarithmically. The low temperature logarithmic variation is caused by the resonant response of coherently driven TLSs in the low frequency limit. At higher temperatures relaxation sets in leading to the minimum and the subsequent increase. As mentioned before, the weak temperature variation of $\varepsilon'(T)$ at the lowest temperatures, sometimes called "dielectric saturation", is not predicted by the TM. Moreover, the theoretically expected ratio of --2:1 of the two log($T$)-slopes is not observed, either. Instead, as in previous experiments \cite{rogge,bechinger}, the ratio is found to be closer to --1:1. 

Within the TM the resonant part of the permittivity is given by 

\begin{equation}
\varepsilon'_{\rm res}-1 = \frac{2\overline{P} p^2}{3\varepsilon_0} \int\limits_{{\it\Delta}_{\rm 0,min}}^{E_{\rm max}}{\rm d} E\, \frac{\sqrt{E^2 - {\it\Delta}_{\rm 0,min}^2}}{E^2} \tanh\left ( \frac{E}{2k_{\rm B}T}\right ),
\end{equation}
and formally, the saturation effect of our sample can be accounted for by a cut-off energy ${\it\Delta}_{\rm 0,min} / k_{\rm B}=12.2\,$mK which is a remarkably high value. The quantity $\overline P p^2/\varepsilon_0$ is determined to be $1.03\times 10^{-2}$, where $p$ is the average magnitude of the dipole moment of the TLSs.

At low frequencies the dielectric loss is caused by relaxation and $\tan \delta \propto T^3$ is expected since one-phonon processes should be dominant. However, as shown in Fig.~1 the measured temperature variation is considerably weaker. It is worth  noting that analogous deviations from the prediction of the TM have been observed in low frequency acoustic experiments.

A further intriguing effect, which cannot be explained by the TM either, is the non-linear dielectric response to the amplitude of the excitation voltage as shown in Fig.~2. With increasing voltage the minimum of $\varepsilon'$ is shifted towards higher temperatures and leads at the same time to different plateau heights at lowest temperatures. The dependence of the dielectric response on the amplitude of the applied ac-field has been treated theoretically by Stockburger {\it et al.} \cite{stockburger}. In contrast to our observation, however, their model predicts a change of the ln($1/T$)-slope and a voltage independent plateau. We suppose that the dependence on the amplitude of the applied ac-field is a direct consequence of electric flux acting on the quantum-mechanical state of TLSs and is therefore linked to the surprising magnetic field effects which will be discussed in the remainder of this paper.

\begin{figure}[b]
\vbox{
\includegraphics[width=0.41\textwidth]{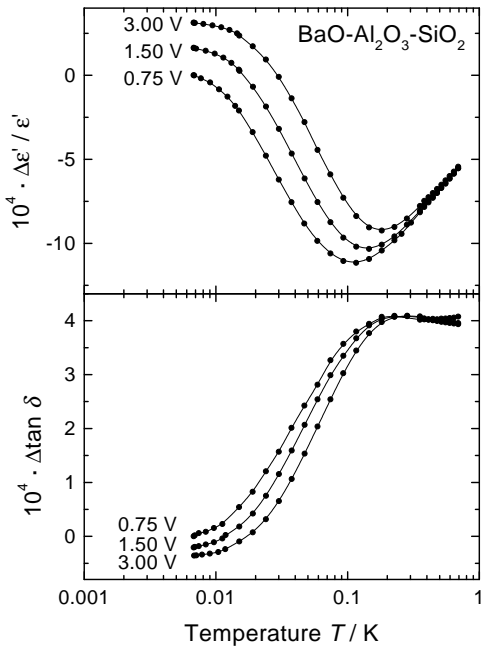}
\caption{Temperature variation of the permittivity $\Delta\varepsilon'/\varepsilon'$ and of the loss $\Delta\tan\delta$ of BaO-Al$_2$O$_3$-SiO$_2$-glass measured at a frequency of $1\,$kHz and excitation voltages $U = 0.75, 1.5,\; {\rm and} \; 3\,$V. The data point at $T_0=1.26\,$mK and $U=0.75\,$V was taken as a reference.}
}%
\label{fig2}
\end{figure}  

Astonishing phenomena were found in magnetic fields. As shown in Fig.~3 the permittivity in the resonant regime varies non-monotonically with the magnetic field: \hbox{$\Delta\varepsilon' (B)/\varepsilon' = [\varepsilon'(B) -\varepsilon' (B=0)]/\varepsilon'(B=0)$} increases with field strength, passes through a first maximum at about $30\,$mT, and exhibits a second one around $250\,$mT. As indicated in the insert of this figure another even more pronounced maximum occurs at the high field of about $18\,$T. In the lower part of the figure the variation of dielectric loss $\Delta\tan\delta=\tan\delta(B)- \tan\delta(B=0)$ is drawn which exhibits an oscillatory behavior, too: With increasing field strength the small rise at low fields is followed by a strong decrease. After going through a minimum the loss increases again. 

\begin{figure}[b]
\vbox{
\includegraphics{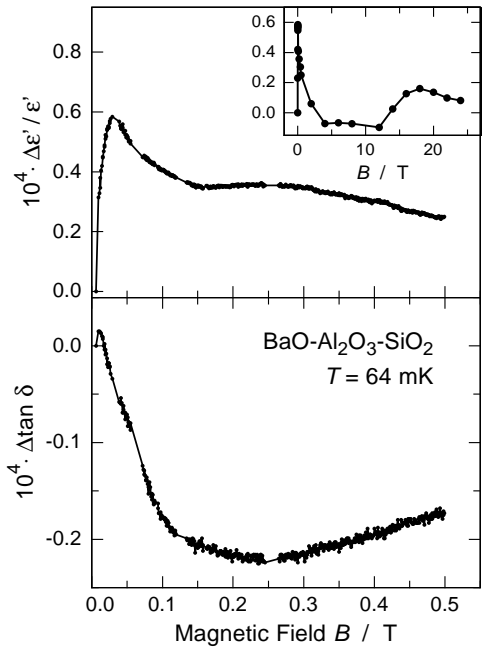}
\caption{Magnetic field dependence of  the permittivity $\Delta\varepsilon' /\varepsilon'$ and the loss angle $\Delta\tan\delta$ of the BaO-Al$_2$O$_3$-SiO$_2$-glass at $T=64\,$mK. The insert shows the behavior at high magnetic fields.} 
}%
\label{fig3}
\end{figure}

An interpretation of the surprising magnetic field dependence of the dielectric response of glasses at low temperature can be given in a generalized tunneling model~\cite{kettemann}. There, the motion of a particle with charge~$Q$ on a closed path with a double-well potential is considered. The presence of an induction field~$\bbox B$ violates the time reversal invariance due to the Aharonov-Bohm phase and leads to an energy spectrum of the TLSs which varies periodically with magnetic flux. The energy splitting of the ground state is then given by $E(\phi) = \sqrt{{\it\Delta}^2+t(\phi)^2}$, where the tunneling splitting $t(\phi)={\it\Delta}_{0} \cos\, (\pi\phi/ \phi_{0})$ depends periodically on the magnetic flux $\phi$ through the area enclosed by the path of the tunneling particle. The period of the oscillation is determined by $\phi_0= h/Q$. The magnetic field causes the TLSs to carry a persistent tunneling current resulting in a magnetic moment. The low-temperature thermodynamic properties such as the specific heat or the permittivity calculated in the generalized model of independent TLSs are consequently periodic functions of the magnetic flux. The resonant part $\varepsilon'_{\rm res}$ of the permittivity can be calculated using Eq.~(1) as before, after substituting the tunneling parameter ${\it\Delta}_0$ and its lower limit ${\it\Delta}_{\rm 0,min}$ by the flux dependent quantities ${\it\Delta}_0 \cos\,(\pi\phi /\phi_0)$ and ${\it\Delta}_{\rm 0,min}\cos\,(\pi\phi/\phi_0)$, respectively. ${\it\Delta}_{\rm 0,min}$ vanishes for $\phi=\phi_0/2$ and $\varepsilon'_{\rm res}$ should exhibit a maximum. In this case the integral in Eq.~(1) can be approximated by $\ln(1/T)$ meaning that $\varepsilon'_{\rm res}(T)$ is expected to vary logarithmically with temperature in accordance with the TM.

\begin{figure}[b]
\vbox{
\includegraphics{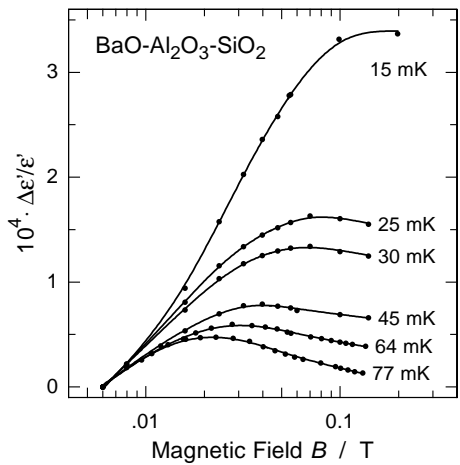}
\caption{Magnetic field variation $\Delta\varepsilon'/\varepsilon' $ of the permittivity  between 15$\,$mK and 77$\,$mK at small magnetic fields.}
}%
\label{fig4}
\end{figure}

The experimental confirmation of this prediction follows from Fig.~4 where the variation of the permittivity with the magnetic field is shown for different temperatures. The maximum values of $\Delta \varepsilon'(B)$ were taken from this graph and plotted in Fig.~1 as open circles. These data points fall approximately onto the curves predicted by the TM for independent TLSs with the small value of ${\it\Delta}_{\rm 0,min} \ll 1\,$mK. It is worth mentioning that already a field of the order of $100\,$mT completely lifts the saturation of $\varepsilon'_ {\rm res}(T)$ at low temperatures. Similarly, the minimum values of $\Delta\tan \delta$ have been taken and drawn in the lower part of Fig.~1. These data points coincide approximately with the \hbox{$T^3$-dependence} predicted by the TM.

As mentioned above, the specific heat is expected to exhibit an oscillatory behavior like the permittivity \cite{kettemann}. Although we did not measure this quantity we have observed that the time needed to reach thermal equilibrium after changing the external magnetic field depends on the applied field. It seems that the variation of $\varepsilon'$ is accompanied by a corresponding changes of the specific heat.

From Fig.~3 we estimate that the oscillation period of ${\varepsilon' (B)}$ is roughly $200\,$mT. Consequently a charge of $Q \approx 4\times 10^5 |e|$ is required, where $e$ is the elementary charge. According to \cite{kettemann} this large value originates from the strong coupling between the TLSs. This is consistent with the large renormalized value of ${\it\Delta}_{\rm 0,min}/k_{\rm B}=12.2\,$~mK which also indicates that the coupling between the TLSs is rather strong. Assuming an average distance between the TLSs of $10^{-8}\,$m and an averaged dipole moment of $p=2|e|\times 10^{-10}\,$m the mean dipole-dipole interaction energy~$U_{\rm int}$ is estimated to be of order $U_{\rm int}/k_{\rm B}\approx 100\,$mK. This means that flux effects should indeed become observable below $T\approx U_{\rm int}/k_{\rm B}\approx 100\,$mK. It has been shown \cite{kettemann} that under certain conditions the excitation spectrum of $N$ strongly coupled TLSs with charge $q$ is equivalent to that of a single particle with charge~$Q = Nq$ tunneling along a closed path. Therefore, it is tempting to introduce quasiparticles whose tunneling path are pierced by a flux with the periodicity $\phi_0=h/(Nq)$.

The theoretical considerations by Kettemann et al. \cite{kettemann} show that another mechanism exists which contributes to changes in the energy spectrum of coupled TLSs, too. The dipole moment of asymmetrical TLSs increases with decreasing tunneling splitting~$t(\phi)$, and thus varies with the external induction field~$\bbox B$. This implies that the dipolar coupling between TLSs also depends on $\bbox B$. Thus, the energy of the ground and the first excited state of a cluster of two or three coupled TLSs may cross already at weak magnetic fields. As a result the value of ${\it\Delta}_0$ and therefore also of ${\it\Delta}_{\rm 0,min}$ is altered by the magnetic field and consequently also the dielectric response of the TLSs.

In a quantitative analysis the average over all orientations and over the charges carried by the clusters of coupled TLSs must be calculated. Thus the oscillations become smeared out especially at higher magnetic fields. In addition, the decrease of the dielectric loss and its oscillatory behavior indicate that the magnetic field has also an influence on the dynamics of coupled TLSs, in particular on their relaxation times. 
In a comprehensive treatment of the phenomena the interplay of the two effects, level crossing and strong coupling between TLSs, has to be taken into account. In this way it should be possible to understand also the maximum in $\varepsilon'$ observed at $18\,$T.

As demonstrated by Fig.~4 the magnetic effects become increasingly pronounced with decreasing temperature. Therefore, the question arises: What happens at ultra-low temperatures? It seems that a transition takes place from mesoscopic to macroscopic large clusters of strongly coupled TLSs. In our view the phase transition reported recently to occur at $5.84\,$mK in BaO-Al$_{2}$O$_{3}$-SiO$_{2}$-glass \cite{strehlow1} is therefore intimately connected with the observations discussed here.

The authors thank G. Schuster, A. Hoffmann, D.~Hechtfischer, and P. van der Linden for helping us with the realization of the experiments. The work has been supported through the DFG (Grant Hu359/11) and the TMR Programme \hbox{(n$^\circ$ ERBFMGECT950077)}.

\end{document}